# Multi-Agent Closed-Loop Reasoning for Organic Structure Elucidation from Multimodal Spectra

Bingsen Xue[#, 1], Zhuojun Jiang[#, 2], Jianhao Zhang[1], Mingcheng Gu[1], Yizhe Yuan[1], Yongtai Zhuo[1], Yifan Zhang[2], Li Wang[2], Ya Su[3], Yue Yuan[4], Jiang Liu[5], Xueqian Kong[6], Cheng Jin*[,1,7,8,9]

[1]Institute of Medical Robotics, School of Biomedical Engineering, Shanghai Jiao Tong University, Shanghai 200240, China
[2]CAS Key Laboratory of Green Process and Engineering, Institute of Process Engineering, Chinese Academy of Sciences, Beijing 100190, China
[3]Department of Neurology, National Centre for Neurological Disorders, National Clinical Research Centre for Aging and Medicine, Huashan Hospital, Fudan University, Shanghai 200240, China
[4]Department of Microbiology Laboratory, Minhang District Center for Disease Control, Shanghai 200240, China
[5]Department of Electrical and Computer Engineering, Johns Hopkins University, Baltimore, USA
[6]School of Chemistry and Chemical Engineering, Shanghai Jiao Tong University, Shanghai 200240, China
[7]Beijing Anding Hospital Capital Medical University, Beijing, China
[8]National Clinical Research Center for Kidney Diseases, Beijing, China
[9]Institute of Digital Medicine, Shanghai Jiao Tong University, Shanghai, China

[#]These authors contributed equally to this work.
[*]Correspondence: chengjin520@sjtu.edu.cn (C.J.).

## Summary:

Following the molecular discovery and synthesis revolutions, scalable automated structure elucidation from routine spectroscopic data remains an outstanding challenge. Despite decades of computational efforts, no existing system achieved reliable reasoning over unseen spectra. Here, we propose MACROS, a multi-agent system automating structure elucidation by emulating expert iterative hypothesis-testing. Trained on 100M simulated and 1.6M experimental spectra-molecule pairs, it natively supports arbitrary combinations of routine spectroscopic techniques. It achieves unprecedented zero-shot generalization to diverse real-world samples, correctly identifying synthetic compounds, natural products and metabolites above 500 Da with 1D NMR. Remarkably, MACROS spontaneously recovers textbook spectroscopic correlations from unassigned data and exhibits emergent chemical intuition such as a ring-first parsing preference, learning fundamental chemical principles rather than memorizing database patterns. MACROS augments chemists via collaboration to deliver sixfold faster, 40% more accurate elucidation. MACROS establishes a scalable foundation for fully automated structure elucidation, and catalyzes accelerated molecular discovery toward autonomous laboratories.

## Introduction

Molecular structure determination is a foundational step in chemical, biological, and biomedical discovery, underpinning natural product research, drug development, studies of biomolecular function, and synthetic chemistry [1–3]. Precisely resolving and quantifying biomolecular structures in the complex intracellular environment is also a fundamental challenge in cell biology, which links phenotype to molecular state and function [4,5]. High-throughput synthesis, automated experimentation, and modern spectroscopy have dramatically expanded the production and structural characterization of small molecules, metabolites, and biomolecular ligands. Yet scalable and fully automated de novo structure elucidation from spectroscopic data remains an outstanding challenge. Even for experienced spectroscopists and conventional computational approaches, molecules with high molecular weight, complex architectures, or dense stereochemistry remain exceptionally difficult to resolve. More fundamentally, the challenge extends beyond automating existing workflows toward building an interpretable elucidation process capable of resolving larger, more complex, and structurally ambiguous molecules with greater precision and robustness[6,7]. Each spectroscopic modality encodes a distinct dimension of molecular information, including local chemical environment, atomic connectivity, stereochemical configuration, and functional-group composition [8,9]. Accurate structure determination therefore requires integrating these complementary signals through sequential, hypothesis-driven reasoning, which has proven difficult to automate [10,11]. Approaches that achieve a step change in both autonomous and interpretable elucidation will thus be transformative for the next generation of biomedical and pharmaceutical discovery.

In real-world laboratory workflows, routine spectroscopic data, including $^{1}$H NMR, $^{13}$C NMR, heteronuclear single quantum coherence (HSQC) NMR, and Fourier-transform infrared (IR) spectra, serve as the primary basis for structural characterization across chemical and biological contexts [12–14]. Mass spectroscopy (MS) and heteronuclear NMR ($^{11}$B, $^{19}$F, $^{31}$P, etc.) can provide unique supplementary information for formula or heteronuclear validation. For reactions or biological samples that produce expected compounds, structural confirmation can be achieved simply by matching spectral peaks against reference standards. By contrast, unexpected or uncharacterized molecules require rigorous de novo structural elucidation. Spectral database retrieval is widely applied to identify molecules with comparable spectral profiles[15,16], yet such methods fail to recognize compounds that lie outside existing libraries. In such cases, only de novo structure elucidation without any library references is feasible, which is fundamentally distinct from library-based retrieval. Human experts perform de novo structure elucidation through iterative hypothesis testing: inferring candidate substructures from chemical shifts and coupling patterns, cross validating these hypotheses across multiple spectral modalities, and refining structural assignments by eliminating inconsistent isomers[17].

Current generative approaches for automating de novo elucidation fall into two categories, yet both remain inadequate for real-world applications. Spectra-guided molecular generation models, when trained on data from specific sources, achieve strong performance on fixed modalities within their training distribution[18,19], but lack the flexibility required to handle real-world spectral variability. Repurposed large language models (LLMs) and vision-language models (VLMs) excel at general chemistry tasks via prompt engineering[20–22], but their performance degrades significantly on structure elucidation [23,24]. Even with massive scale and pretraining on extensive chemical and spectral corpora, these models are fundamentally designed for linguistic or visual-semantic reasoning, not spectroscopic analysis. Structure elucidation from multimodal spectra, by contrast, is a highly symbolic, logic-driven task that relies on spectral-molecular pattern matching, multi-step deduction, and iterative validation [25]. Crucially, neither approach incorporates the explicit iterative reasoning workflow that human spectroscopists employ for complex spectra: hypothesis generation, cross-validation across modalities, and step-by-step refinement. Effective solutions therefore require generative foundation models designed from the ground up for spectroscopic reasoning: (i) self-supervised pretraining on large, unpaired spectral corpora to acquire intrinsic physicochemical knowledge, (ii) inherently flexible architectures that tolerate arbitrary or missing modalities, and (iii) explicit closed-loop reasoning mechanisms that emulate the iterative hypothesis-testing workflow of expert chemists.

Here, we propose MACROS (Multi-Agent Closed-Loop Reasoning for Organic Structure), a purpose-built multi-agent system for multimodal spectroscopic structure elucidation from routine spectra: $^{1}$H NMR, $^{13}$C NMR, HSQC, and IR. Unlike repurposed general-purpose foundation models, MACROS was designed from first principles for spectroscopic reasoning. Modality-specific agents were pretrained in a self-supervised manner on large-scale simulated spectra, integrated into a hierarchical framework and jointly trained on over 100 million spectra-structure pairs, then fine-tuned on unassigned experimental spectra to adapt to real-world spectral variability. At its core, MACROS implements an explicit multi-agent closed-loop reasoning mechanism that emulates the iterative workflow of expert spectroscopists. Spec2Mol proposes candidate structures, Mol2Spec simulates spectra and ranks candidates by multi-criterion similarity, while PromptSpec2Mol refines candidates over multiple cycles using high-ranking outputs as guidance. This closed-loop scheme enables test-time performance scaling, allowing progressive accuracy gains during inference. The flexible agent workflow readily accommodates additional information sources (e.g., reactant structures) and facilitates future extensions to higher-dimensional NMR and other spectroscopic modalities. For instance, MS-derived formulas or fragments, annotated by specialized tools (e.g., SIRIUS[26], BUDDY[27]), are accepted as molecular prompts for indirect incorporation, leveraging mature MS solutions and circumventing the scarcity of paired MS-NMR datasets. Notably, attention-based interpretation recovers textbook-like chemical shift–functional group relationships (e.g., aromatic protons at $^{1}$H 6.5–8.5 ppm, carbonyl carbons at $^{13}$C 195–225 ppm) directly from unassigned spectra, demonstrating that the model adheres to established spectroscopic principles. Without explicit structural rules encoded, MACROS exhibits emergent chemical intuition, exemplified by a ring-first preference for rigid cyclic scaffolds at a rate exceeding 40%. MACROS demonstrated strong zero-shot generalization across diverse real-world molecular classes. By fine-tuning toward domain or structural prior, MACROS yields specialized performance for distinct chemical and biological fields, including synthetic compounds, complex natural products, and human metabolites. Atom-level confidence scores along reasoning trajectories supported interpretable visualization of molecular structures, enabling intuitive human-AI collaboration. MACROS establishes a scalable foundation for fully automated spectroscopic structure elucidation and catalyzes accelerated molecular discovery toward autonomous research.

## Results

### MACROS generalizes across diverse sources for de novo structure elucidation

Developing from multi-source datasets of >100 million spectrum-molecule pairs (**Extended Data Figure 1a**), MACROS demonstrated high accuracy in molecular parsing for arbitrary combinations of routine spectra ($^{1}$H, $^{13}$C, HSQC NMR and IR) without retrieval and any molecular priors (e.g. formulas, reactants) **(Figure 1)**. On randomly selected test set (n=560) from USPTO dataset[28], MACROS was able to predict accurate molecular structure with 0.88 (95% CI 0.86-0.89) fingerprint similarity and over 0.95 recall of common functional groups **(Figure 1b-d)**. Specifically, we observed high improvement of accuracy after the closed-loop process of rethinking and refining (**Figure** 1**b**). Rethinking scores strongly correlate with ground-truth molecular similarity. The combined rethinking score (IR spectral similarity and clipped molecular confidence) yields a concordance index (c-index) of 0.71, correctly ranking the true structure within the top 25% in 52% of cases and within the top 10% in 31% of cases (**Figure 1c**). MACROS was robust to noisy and incomplete spectra (**Extended Data Figure 1b)** and scalable to molecules with larger heavy-atom number than training data. Finally, the model can provide both atom-level and molecular-level confidence scores for human inspection and model rethinking (**Extended Data Figure 1c**).

The high accuracy achieved on the USPTO-derived test set extends consistently to spectra from independent sources (**Extended Data Figure 1d–g**). On the QM9S dataset[29], MACROS achieved a fingerprint similarity of 0.55 (95% CI 0.53–0.57) for the initial draft and 0.71 (95% CI 0.69–0.74) after closed-loop refinement and selection. Comparable performance was observed on the large-scale simulated SimPubChem dataset[30] (0.57 (95% CI 0.55–0.59) initial; 0.67 (95% CI 0.65–0.69) refined). All results reported below were obtained under strict zero-shot conditions (no test-time fine-tuning or modality-specific adaptation; **Extended Data Fig. 1e–f**). On the NMRMIND dataset[31], MACROS attained a fingerprint similarity of 0.47 (95% CI 0.45–0.49) for draft, rising to 0.56 (95% CI 0.54–0.58) after closed-loop refinement. On the NMRGym dataset[32], MACROS achieved a fingerprint similarity of 0.47 (95% CI 0.45–0.49) for draft, rising to 0.64 (95% CI 0.62–0.67) after closed-loop refinement. These results confirmed that the accuracy and robustness of MACROS generalize effectively from the primary training distribution to diverse sources.

**Closed-loop multi-agent reasoning enhances structure elucidation at inference**

MACROS achieved test-time scaling and greatly improved the accuracy of molecular structure prediction, by integrating a hierarchical multi-agent architecture with iterative closed-loop reasoning (**Figure 1e, Extended Data Figure 2a**). Specifically, we developed an iterative closed-loop reasoning process within a hierarchical multi-agent framework, which dynamically coordinated multiple agents to execute specialized tasks, including drafting, rethinking, and refining molecular structures These tasks were seamlessly interconnected through a feedback loop: drafting generated initial SMILES-based molecular structures, rethinking evaluated their validity by simulating spectra and comparing them to experimental data, and refining optimized these structures using similarity scores and molecular confidence metrics. This iterative cycle drove test-time scaling, where deciphering performance improved with increasing iterations. This multi-agent system was structured bottom-up, progressing from modality-specific agents to function-specific subsystems **(Figure 2a)**.

Self-supervised pretraining endowed modality-specific agents with deep chemical and spectral knowledge. Four modality-specific decoder-only transformers were pretrained on over 100 million multi-source spectra via next-token prediction, acquiring universal spectral grammar without labels (**Figure 2b, Extended Data Figure 2b**). Vibrational spectra were tokenized as patches, NMR as discrete peaks, and molecules as SMILES characters (**Figure 2b**). Pretrained solely on spectral sequences, MACROS agents spontaneously discovered vibrational modes, carbon hybridization states, and proton environments, revealing an AI-native chemical intuition emergent from self-supervised learning. Unsupervised clustering of IR agent embeddings reveals strong separation by functional group chemistry, including patterns consistent with O–H, N–H, C–X, N=O, C=O, and C≡C vibrations (**Figure 2c**). Unsupervised clustering of $^{13}$C NMR agent embeddings revealed separation by carbon hybridization ($sp^2$, $sp^3$; **Figure 2d-e**). Similarly, the $^{1}$H NMR agent stratifies proton environments with high chemical homogeneity (**Figure 2f-g**).

Supervised fine-tuning assembled pretrained agents into task-specific subsystems (**Figure 2e-f**). In forward structure prediction, a summary agent integrated spectral encodings and feeds them to a molecular agent that autoregressively generates initial SMILES drafts (drafting workflow). The refining subsystem extended the drafting workflow by incorporating molecular context, e.g. draft molecules, fragments, or reaction precursors (**Figure 2e**). Ablation of modality masking ratios revealed 60% as optimal (**Extended Data Figure 3a**). At this level, MACROS achieved peak performance on both complete and incomplete spectral inputs. Token-level confidence (maximum probability per generated token) increased progressively over the first eight tokens and stabilized thereafter, consistent across low-, medium-, and high-complexity molecules (**Extended Data Figure 3b-c**). This pattern was highly consistent with SMILES syntax ambiguity, as organic molecules had multiple valid starting points and non-canonical forms. Monte Carlo hybrid search was thereby designed to explore the diverse SMILES representation space and outperforms beam and greedy decoding (**Extended Data Figure 3d–g**). Clip confidence (product of post-8th-token probabilities) correlated with Tanimoto similarity (Pearson $r > 0.4$; **Extended Data Figure 3h, i**), enabling high-confidence molecular-level filtering of predicted structures.

In the MACROS rethinking subsystem, a summary agent aggregated encoded molecular tokens and transfers them to spectral agents for spectra simulation, followed by similarity scoring against experimental data (Figure 2d). Infrared similarity was computed directly via cosine similarity. For NMR modalities, peaks were first aligned using the Hungarian algorithm on absolute chemical-shift differences, then normalized to yield a bounded similarity score (**Materials and methods**). In contrast to conventional spectral simulation methods, which rely on computationally intensive 3D conformer generation and Merck molecular force field (MMFF) optimization (typically requiring seconds to minutes per molecule), Mol2Spec predicted spectra directly from SMILES strings in a single forward pass. Furthermore, Mol2Spec can be trained on raw spectrum–structure pairs without requiring peak assignments, thereby substantially reducing the annotation burden and facilitating direct fine-tuning on experimental laboratory data (**Supplementary Note 5**). The spectral simulation agents achieved accurate reverse prediction with low errors ($n = 1000$): $^{1}$H NMR (MAE 0.16 ppm, 95% CI 0.15–0.17 ppm), $^{13}$C NMR (1.34 ppm, 95% CI 1.20–1.48 ppm), HSQC ($^{1}$H: 0.32 ppm, 95% CI 0.30–0.34 ppm; $^{13}$C: 2.29 ppm, 95% CI 2.08–2.50 ppm), and IR (cosine similarity 0.761, 95% CI 0.754–0.768). On IR prediction, these values outperformed DetaNet[29] (0.719, 95% CI 0.712–0.726) and AttentiveFP[33] (0.720, 95% CI 0.713–0.727) (**Extended Data Figures 4, 5a–c**; representative predictions in **Extended Data Figures 4e–g and 5d–f**). Clip token-level probabilities were aggregated into molecular confidence scores, enabling high-confidence structure selection (**Figure 2f**). IR similarity score and clip confidence both correlated strongly with true molecular similarity (**Extended Data Figure 1b**). This closed-loop integration of drafting, rethinking, and refining ensured robust, experimentally aligned structure prediction.

**MACROS learns chemical intuition and interpretable reasoning**

MACROS exhibited interpretable decision-making across system and submodule levels. Although no explicit modality biases were introduced during training, the model preferentially used NMR spectra for forward drafting and IR spectra for rethinking validation. NMR noise impaired both draft generation and final prediction accuracy, whereas IR noise selectively degrades performance in the rethinking validation phase (**Extended Figure 1a**). IR similarity scores correlated strongly with Tanimoto similarity, while NMR scores remained consistently high (**Extended Figure 1b**). We hypothesized that MACROS leverages NMR chemical shift information to produce robust draft predictions with high NMR-based structural similarity, while IR spectra, alongside clip-confidence metrics, facilitate validation in the rethinking phase. Modality ablations confirmed this asymmetry: excluding IR causes minimal performance drop, whereas excluding NMR severely impairs accuracy (**Figure 1b**).

To understand the internal reasoning process of the model, we analyzed attention maps at the final SMILES generation step, aggregating scores for tokens corresponding to 13 functional groups **(Figure 3a, Extended Data Figure 6)**. MACROS recovered spectroscopic consensus, e.g., ketones ($^{1}$H: 2–3 ppm; $^{13}$C: 195–225 ppm), aldehydes ($^{1}$H: 9–10 ppm; $^{13}$C: 190–205 ppm), without peak-level supervision **(Figures 3b–c)**. The same mechanism in rethinking maps predicted spectra back to molecular substructures, yielding consistent, interpretable alignments (**Extended Figures 4a, c**). It was worth noting that the chemical-shift/functional-group relationships compiled in this study originated from decades of empirical practice, and their shift accuracy remained contingent on factors such as spectrometer configuration, choice of deuterated solvent, and measurement conditions. Consequently, the interpretation precision of MACROS in assigning functional-group-specific chemical shifts was fundamentally related to the accuracy and consistency of the underlying databases. This close correspondence further confirmed the reliability of the data source. These findings demonstrated that MACROS acquires expert-like chemical intuition through self-supervised spectral modeling, enabling reliable, automated structure elucidation. Crucially, no explicit correspondence between individual spectral peaks and molecular substructures was ever provided during training. The sole supervisory signal consisted of raw multimodal spectra paired with correct SMILES. Nevertheless, following large-scale self-supervised pretraining on unpaired spectral data and subsequent end-to-end supervised fine-tuning on annotated pairs, MACROS autonomously recovered the entire canonical chemical-shift–functional-group correlation table with textbook fidelity.

Predictive patterns exhibited a consistent and chemically intuitive strategy of structure elucidation. Across diverse datasets (QM9, NMRMIND, USPTO, and SimPubchem), MACROS consistently initiated SMILES sequences with ring systems (>40%, **Figure 3d**). This ring-first preference directly mirrored the expert heuristic of anchoring interpretation on rigid cyclic scaffolds before appending substituents, despite the autoregressive decoder receiving no explicit structural or syntactic priority during training[16,17]. During training phase, usage of non-canonical SMILES can enhance the priority (**Figure 3d**). UMAP visualizations of the learned $^{1}H$ and $^{13}C$ NMR chemical-shift embeddings from the pretrained NMR agents provided direct mechanistic support for this intuition (**Figure 3d**). Trajectories exhibited pronounced directional changes near the diagnostically critical regions of $^{13}C \approx 120$ ppm and $^{1}H \approx 7$ ppm, regions classically associated with aromatic systems (**Extended Data Figure 7**). This pattern emerged already during self-supervised pretraining on unpaired spectra and was subsequently refined during supervised fine-tuning, indicating that the model autonomously identifies aromatic motifs as highly salient features. The observed ring-first generation strategy thus reflected not only the acquisition of core spectroscopic–structural correlations but also the emergence of an expert-like hierarchical reasoning process that systematically privileges rigid cyclic fragments as primary interpretive anchors.

**MACROS generalizes to real-world spectra and augments human experts**

To evaluate zero-shot performance on authentic experimental spectra, we assembled a curated real-world evaluation set of 1,701 molecules from multiple independent sources, including randomly sampled entries from NMR-BANK (n = 540), literature-reported spectra from recent J. Am. Chem. Soc. publications (2019–2024; n = 450), challenging natural products with HSQC data (n = 398), and a subset from the NMR-Exp database[24] (n = 313). All structures were verified independently and confirmed to be absent from the model's training and finetuning corpus.

**Validations on natural products.** A randomly selected subset of the natural-product cohort from NP-MRD was used for evaluation (n = 380 unique compounds, each with experimental $^{1}H$ and $^{13}C$ NMR data; mean heavy atom count 31.2, 95% CI: 29.8–32.7, range 8–112; mean molecular weight 438 Da, 95% CI: 417–458, range 112–1570). On this subset, the zero-shot prediction achieved an initial draft fingerprint similarity of 0.559 (95% CI: 0.535–0.585). This increased markedly to 0.625 (95% CI: 0.600–0.651) after closed-loop reasoning (**Figure 4a**). Targeted fine-tuning on natural products further elevated performance to 0.799 (95% CI: 0.778–0.820), with an additional improvement to 0.843 (95% CI: 0.822–0.864) following reasoning (**Figure 4b**). In total, 54.2% of the molecules were correctly resolved from the spectra. Zero-shot predictions successfully captured broad structural diversity, including terpenoids, prenylated flavones, flavonoid glycosides, and lignan glycosides **(Figure 4i, Extended Data Figure 9**). Moreover, MACROS demonstrated promise for stereoisomer discrimination, encompassing enantiomers and cis–trans isomers (**Extended Data Figure 10**). Notably, MACROS initially trained on generic organic compounds achieved high-fidelity structural assignment of diverse natural products after only minimal fine-tuning with partial domain-specific knowledge, highlighting efficient transfer learning that avoids the need for extensive specialized training.

**Validations on literature-reported spectra.** A randomly selected subset of the natural-product cohort from NMRBANK comprised 544 unique samples with experimental $^{1}$H and $^{13}$C NMR data (mean heavy atom count: 25.6, 95% CI 24.9–26.4, range 3–62; mean molecular weight: 369 Da, 95% CI 358–379, range 69–970 Da). On this subset, zero-shot predictions yielded a Tanimoto similarity of 0.478 (95% CI 0.460-0.497) (**Figure 4c**, **Extended Data Figure 11**). Fine-tuning on data of chemical compounds raised performance to 0.742 (95% CI 0.705-0.779), demonstrating rapid adaptation to literature-derived spectral distributions. A randomly selected subset of the natural-product cohort from NMRExp comprised 313 unique samples with experimental $^{1}$H and $^{13}$C NMR data (mean heavy atom count: 24.9, 95% CI 23.8–25.9, range 9–101; mean molecular weight: 355 Da, 95% CI 340–369, range 119–1414 Da). On this subset, zero-shot performance yielded an initial draft similarity of 0.431 (95% CI 0.405,0.457), improving to 0.479 (95% CI 0.452-0.507) after closed-loop refinement. After fine-tuning, draft similarity increased to 0.683 (95% CI 0.649-0.711), reaching 0.743(95% CI 0.714-0.772) after reasoning. On in-house acquired crude reaction products using $^{13}$C NMR alone, MACROS attained a median similarity of 0.456 (95% CI 0.440–0.472) zero-shot (**Supplementary Note 8**), with robust prediction despite severe signal overlap and impurities. These results establish that MACROS delivers robust zero-shot capability across diverse real-world data and can be rapidly specialized to laboratory-specific chemical spaces with minimal additional finetuning.

**Validations on metabolite molecules.** The metabolic cohort from BMRB comprised 409 unique samples (mean heavy atom count: 18.7, 95% CI 17.5–19.9, range 2–76; mean molecular weight: 268 Da, 95% CI 251–284, range 32–1062 Da). On this cohort, using only $^{1}$H and $^{13}$C NMR data, the zero-shot prediction achieved an initial draft similarity of 0.519 (95% CI: 0.491-0.546). It improved markedly to 0.589 (95% CI: 0.560-0.618) following closed-loop reasoning (**Figure 4e**). After fine-tuning, performance rose substantially to 0.742 (95% CI: 0.713,0.770), and further increasied to 0.816 (95% CI: 0.790,0.843) after reasoning (**Figure 4f**). In total, 68.9% of the molecules were correctly resolved from the metabolic spectra. The zero-shot predictions successfully encompassed a broad diversity of metabolic classes, including amino acids, alkaloids, nucleotides, and glycosides (**Figure 4j**). Despite decades of metabolomics research, known metabolites represent only a tiny fraction of possible structures in the mammalian metabolome, leaving extensive unidentified signals as metabolic “dark matter”. MACROS potentially enables proactive discovery of unknown metabolites from NMR data, providing a complementary alternative to DeepMet's passive MS-based validation strategy and supporting systematic exploration of metabolic dark matter in NMR-centric datasets[34,35].

**Validations on autonomous reaction annotations.** Spectral analysis of reaction products remains a cornerstone of mechanistic elucidation and real-time reaction monitoring in synthetic chemistry. MACROS introduces reactant structures as molecular prompts to the PromptSpec2Mol subsystem, enabling substantial accuracy improvements without any explicit training on reaction mechanisms or outcome prediction. On the USPTO benchmark (n = 360), performance with reactant prompting yielded an average TFPS 0.844 (95% CI 0.825–0.863) for the initial draft, rising to 0.889 (95% CI 0.874–0.904) after rethinking. MACROS was further evaluated on 450 real-world crude reaction products curated from recent J. Am. Chem. Soc. publications (2019–2024). In total, MACROS successfully annotated 44% of chemical reactions from real-world cases using zero-shot prediction. Using only $^{13}C$ NMR spectra combined with reactant molecular prompts, the zero-shot prediction achieved a draft TFPS of 0.556 (95% CI 0.533–0.580), increasing to 0.5886 (95% CI 0.566–0.612) after rethinking (**Figure 4k**). Using $^{1}H$ and $^{13}C$ NMR spectra combined with reactant molecular prompts, the zero-shot prediction achieved a draft TFPS of 0.642 (95% CI 0.620–0.664), increasing to 0.677 (95% CI 0.656–0.698) after rethinking. Selected cases were demonstrated in **Extended Data Figure 12 and Supplementary Note 9.** These demonstrate that contextual prior information from reactants alone can effectively constrain the combinatorial search space and steer predictions toward chemically plausible products.

**Human–AI collaboration in real-world spectral analysis.** MACROS enables efficient characterization of complex chemical systems in automated laboratory workflows and supports chemists in precise product elucidation. In a curated PubChem dataset (chemical engineering, environmental, pharmaceutical molecules), MACROS alone achieves >90% structural accuracy in <1 minutes per molecule on a single 24 GB GPU. A blinded human-expert study was conducted with three organic spectroscopists (**Materials and Methods**). Each expert first elucidated structures from raw multimodal spectra without assistance (about 1 h per molecule, 0.102 TFPS). After a four-week wash-out period, the same spectra were re-analyzed with MACROS assistance (full closed-loop mode with confidence scores displayed). Mean time was reduced to about 10 min per molecule and TFPS increased to 0.807 (**Figure 4l**). Closed-loop reasoning trajectories further reveal systematic refinement (**Figure 4l**). Initial drafts exhibit high structural diversity; each iteration prunes low-confidence candidates, monotonically increasing molecular confidence, reverse spectral similarity, and ground-truth TFPS. This convergence is most pronounced for complex natural products and macrocycles, where accuracy gains exceed 60% within five cycles. These results demonstrate that MACROS not only enables fully autonomous spectral analysis at expert level, but also dramatically accelerates and augments human expertise, establishing a practical symbiosis between chemists and AI in modern automated laboratories.

## Discussion

Spectral-to-structure elucidation remains one of the most cognitively demanding tasks in organic chemistry, requiring iterative hypothesis testing across complementary yet heterogeneous modalities. Chemical experts require considerable breadth of thought when interpreting spectra. Typically, this involves dissecting spectral peaks and the complex relational logic of molecular structures from multiple perspectives, ultimately identifying a molecular structure that corresponds to all observed peaks. In this process, the presence of shifted or interfering peaks introduced by complex systems further amplifies the difficulty of this "logical puzzle". Elucidating molecular structures from spectroscopic data is a complex, systematic task that necessitates the integration of multidisciplinary knowledge. While the structural information afforded by routine spectroscopic methods is insufficient for unambiguous, full molecular elucidation, especially for highly complex species, these approaches remain indispensable for exploring the vast expanse of uncharted chemical space. For automated molecular structure elucidation, the key is to establish a scalable foundation with correct spectroscopic principle, which can be readily extended to novel modalities and specialized fields.

Here, we propose MACROS, a hierarchical multi-agent framework that emulates the iterative reasoning of expert spectroscopists. MACROS achieves robust performance on arbitrary combinations of routine spectra and exhibits robust zero-shot generalization to diverse real-world experimental spectra, including chemical reactions, natural products and metabolites. On a single consumer-grade GPU (24 GB), MACROS processes >20 molecules per minute in quick mode and >1 molecules per minute in full closed-loop mode. MACROS streamlines expert-driven structure elucidation, delivering sixfold faster analysis and substantially enhanced accuracy. Through large-scale pretraining and closed-loop reasoning, MACROS establishes a scalable foundation for generalizable molecular structure elucidation from multimodal spectra, with the potential to support self-driving laboratories for molecular discovery and validation.

This advantage derives in part from deliberate modality-specific pretraining on native data representations. Rather than forcing heterogeneous spectroscopic data into generic text or image representations, we pretrain lightweight agents directly on their physically native formats that peak lists for NMR, continuous waveforms for IR, and chemistry-specific tokenization for SMILES. This preserves the intrinsic inductive biases that are inevitably lost in monolithic LLM/VLM backbones. The resulting agents are integrated through lightweight projection layers into task-specific subsystems, achieving markedly higher accuracy than contemporary LLM-based approaches while requiring orders-of-magnitude fewer parameters for adaptation. Leveraging multi-source large-scale self-supervised pretraining, individual agents within MACROS acquire broadly generalizable spectroscopic and molecular knowledge. This positions them as promising candidates for serving as foundational models tailored to NMR, IR, and molecular representation learning more broadly, thereby enabling robust multimodal molecular networking [42,43] . This physically grounded, reusable agent paradigm offers a practical, resource-efficient alternative to the prevailing strategy for scientific data analysis.

Equally important is the closed-loop reasoning framework, which enables test-time scaling without gradient updates. Unlike approaches that rely on reinforcement learning to discover elucidation strategies autonomously, MACROS employs an explicit multi-agent workflow whose interaction patterns are deliberately derived from expert organic chemists' systematic reasoning. The rethinking stage provides a chemistry-native consistency check from an orthogonal physical perspective: candidate structures are re-encoded and passed through reverse spectral agents, yielding predicted NMR chemical shifts and IR band patterns that are scored against raw experimental data. This step effectively measures agreement between hypothesis and observation using physically meaningful distances rather than learned proxies. The refining stage further extends the base drafting model through targeted fine-tuning on prefix-conditioned inputs. By feeding spectroscopically validated fragments (reactants or top drafts) as prefixes, the generation is restricted to the low-dimensional detail subspace around chemically correct anchors, transforming exponential de novo search into efficient local optimization while preserving established structural integrity. Although the current agent topology is deliberately minimal and fully hand-crafted for chemical transparency, it already delivers expert-level spectral parsing across highly diverse real-world datasets. More refined workflow design in future iterations could allow MACROS to tackle an even broader range of molecular discovery tasks with the same transparent, expert-inspired closed-loop principle.

Interpretability analyses confirm that MACROS faithfully recapitulates the canonical correspondence between spectral features and molecular substructures. Modality ablation and noise-perturbation studies reveal a learned diagnostic hierarchy that mirrors expert practice. Attention scores further demonstrate that MACROS autonomously reconstructs the canonical mapping between NMR chemical shifts and molecular substructures. Remarkably, MACROS has never shown any explicit correspondence between chemical shifts and molecular structures, yet it rediscovered these empirical, textbook-level correlations purely from raw paired data, illustrating a scalable paradigm for extracting latent natural laws directly from unannotated experimental records. This emergent rediscovery of textbook spectroscopic–structural correlations, traditionally distilled over decades of chemical experimentation, illustrates a broader principle. This capability establishes a transferable framework to mine latent principles from unannotated experimental data, with broad potential for de novo multi-modal data interpretation. MACROS displays emergent chemical intuition through a ring-first structural assembly strategy, preferentially initiating SMILES generation with rigid cyclic scaffolds across all tested datasets. This data-driven, AI-native understanding of structure elucidation aligns with the heuristic preferences refined by expert chemists through extensive practical experience. Collectively, these findings confirm that MACROS has initially developed an expert-like reasoning framework for structure elucidation.

MACROS represents a significant step forward in autonomous chemical research, offering an efficient and stable method for molecular structure elucidation through its multi-agent, closed-loop architecture. By decoupling modalities and leveraging domain-specific knowledge, MACROS outperforms general-purpose LLMs while maintaining computational efficiency. Its explicit multi-agent design and test-time scaling capabilities provide a flexible framework for addressing complex scientific challenges. Future evolution, including broader datasets, improved training and inference strategies, and tighter integration with automated experimentation, should further expand its utility in chemical synthesis, natural product discovery and automated laboratories. This work underscores the potential of specialized AI systems to transform scientific discovery, offering a blueprint for future advancements in computational chemistry and beyond.

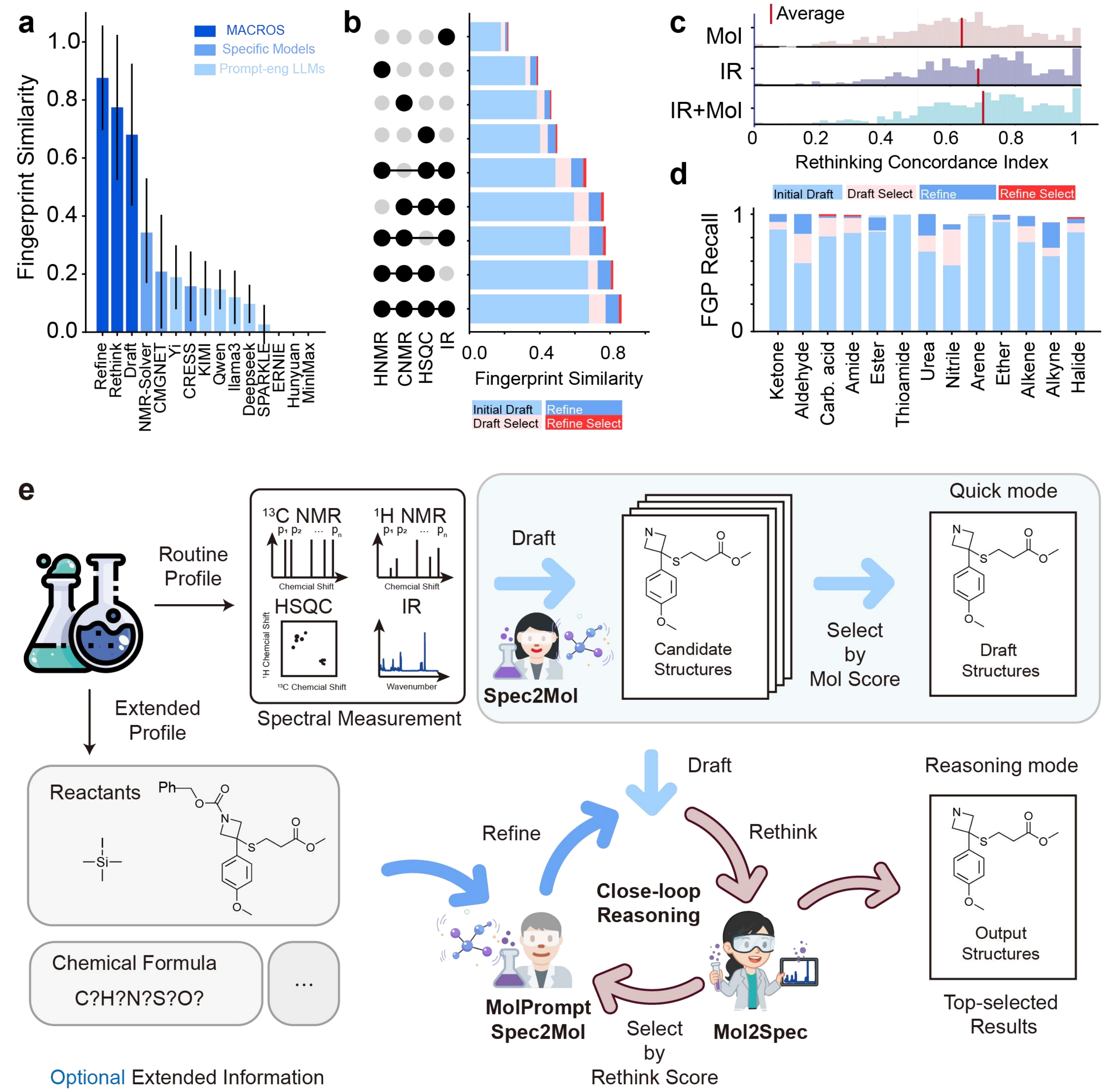

**Figure 1. Multi-agent systems for closed-loop multi-modal spectral reasoning.**

**a.** Performance benchmark on the USPTO-derived test set (n = 200 compounds). Fingerprint similarity (± 95% confidence interval) of predicted versus ground-truth structures is shown for MACROS and current state-of-the-art methods, including dedicated spectrum-to-structure models and prompt-engineered large language models (LLMs). **b.** Accuracy scaling with reasoning for different multimodal input combinations ($^1$H NMR, $^{13}$C NMR, HSQC, IR). Performance of the initial draft (iteration 0) and subsequent refinement cycles is reported. **c.** Concordance between candidate ranking by rethinking scores (IR spectral similarity, clipped molecular confidence, and their weighted combination) and ranking by ground-truth fingerprint similarity across the full test set. **d.** Functional-group recall of initial drafts versus final closed-loop predictions for 13 common organic functional groups. **e** Closed-loop reasoning workflow for test-time scaling in multimodal spectral structure elucidation. The iterative pipeline consists of (i) draft candidate generation using hybrid decoding and uncertainty-aware confidence estimation, (ii) spectral simulation of candidates followed by multi-criteria ranking, and (iii) guided refinement over multiple cycles using top-ranked structures as prefixes. Optional extended information may be incorporated into the workflow, such as reactants, molecular formulae derived from MS, or data from elemental analysis.

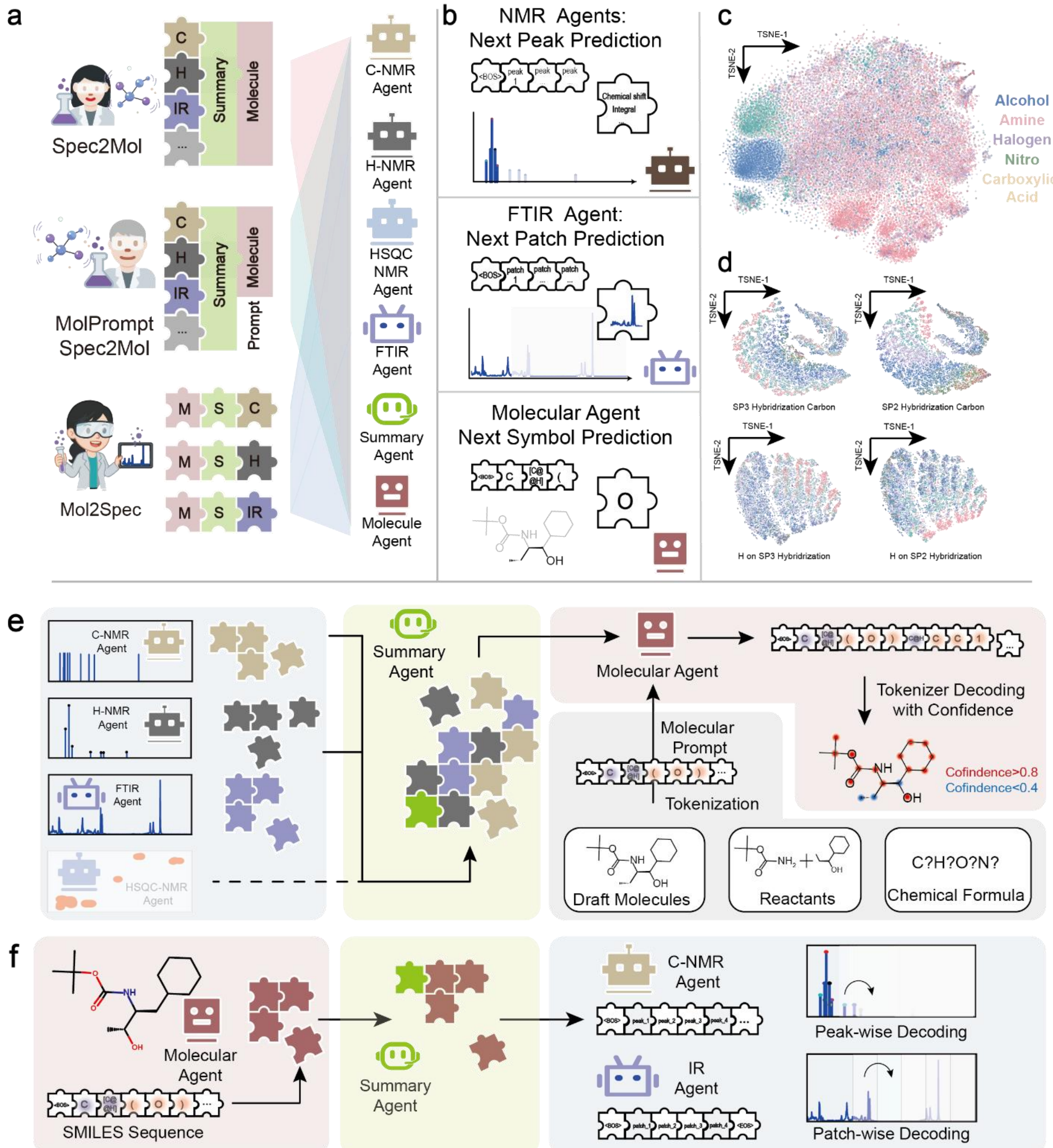

**Figure 2. MACROS development via self-supervised pre-training and fine-tuning.**
**a.** Hierarchical multi-agent assembly. Modality-specific agents ($^{1}$H NMR, $^{13}$C NMR, HSQC, IR, and molecule) are first pretrained independently, then jointly fine-tuned into task-oriented subsystems (forward Spec2Mol and reverse Mol2Spec), and finally integrated into closed-loop workflows that enable iterative drafting, rethinking, and refinement. **b.** Self-supervised autoregressive pretraining paradigm. Each modality-specific agent is pretrained on >$10^7$ unpaired examples of its native data type (peak lists for NMR, full waveforms for IR, SMILES strings for molecules) using a next-token prediction objective. **c-d.** Emergent chemical knowledge from pretraining. Without explicit supervision, the IR agent learns canonical functional-group wavenumber ranges (**c**), the $^{13}$C NMR agent distinguishes sp, sp$^2$, sp$^3$, and quaternary carbons, and the $^{1}$H NMR agent resolves proton environments (representative attention-derived correlations shown) (**d**). **e.** Forward deciphering subsystem (Spec2Mol). Spectral agents encode multimodal inputs; a summary agent integrates features and feeds the base molecular agent for de novo SMILES generation. Optional draft-molecule prefixes can be supplied for guided refinement. **f.** Reverse spectral simulation subsystem (Mol2Spec). The large molecular agent encodes candidate structures; spectral agents independently predict corresponding $^{1}$H NMR, $^{13}$C NMR, HSQC, and IR data, enabling similarity-based candidate scoring in the rethinking phase.

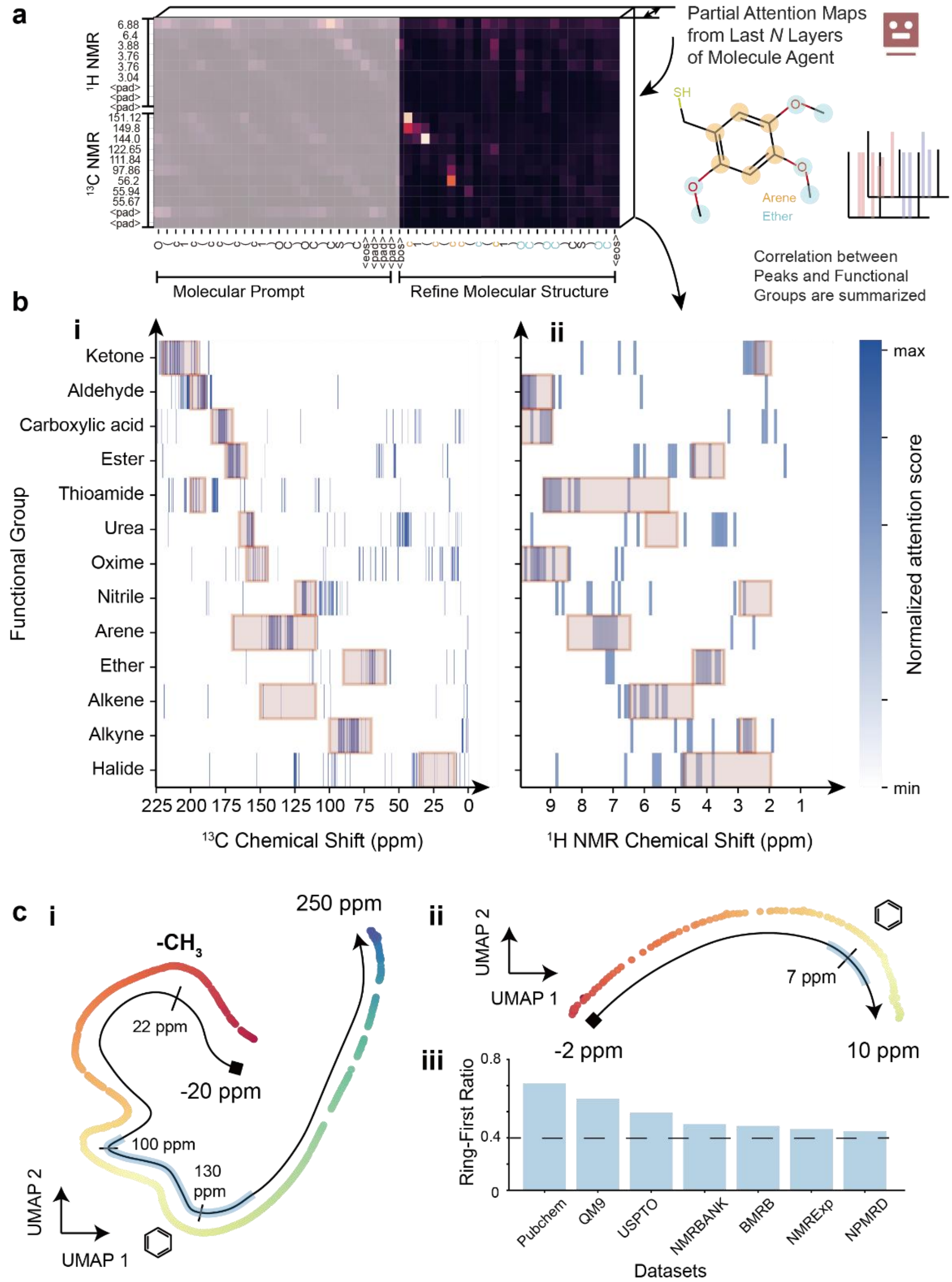

**Figure 3. Interpretation of spectral parsing process at multi-scale.**

**a.** Attention-based interpretability. Causal attention weights from the final three transformer layers of the molecular agent are pruned to the top 10% per query token, averaged across heads, and aggregated by predefined functional groups. Heatmaps show attention from SMILES tokens (x-axis) to spectral peaks (y-axis).

**b.** Learned $^{1}$H (**i**) and $^{13}$C (**ii**) NMR chemical-shift–functional-group correlations. Heatmap shows aggregated attention from functional-group atoms to $^{1}$H NMR peak tokens. Grey shaded regions indicate empirical ranges compiled from decades of experimental NMR literatures [50,51]. Learned correlations highly overlap with empirical ranges

**c.** UMAP projection of learned $^{1}$H (**i**) and $^{13}$C (**ii**) NMR chemical-shift embeddings from the pretrained NMR agents. Trajectories show pronounced directional changes near $^{13}$C ≈ 120 ppm and $^{1}$H ≈ 7 ppm, regions classically associated with aromatic systems. Bar charts (**iii**) confirm that MACROS consistently initiates correct top-1 SMILES predictions with ring systems across datasets.

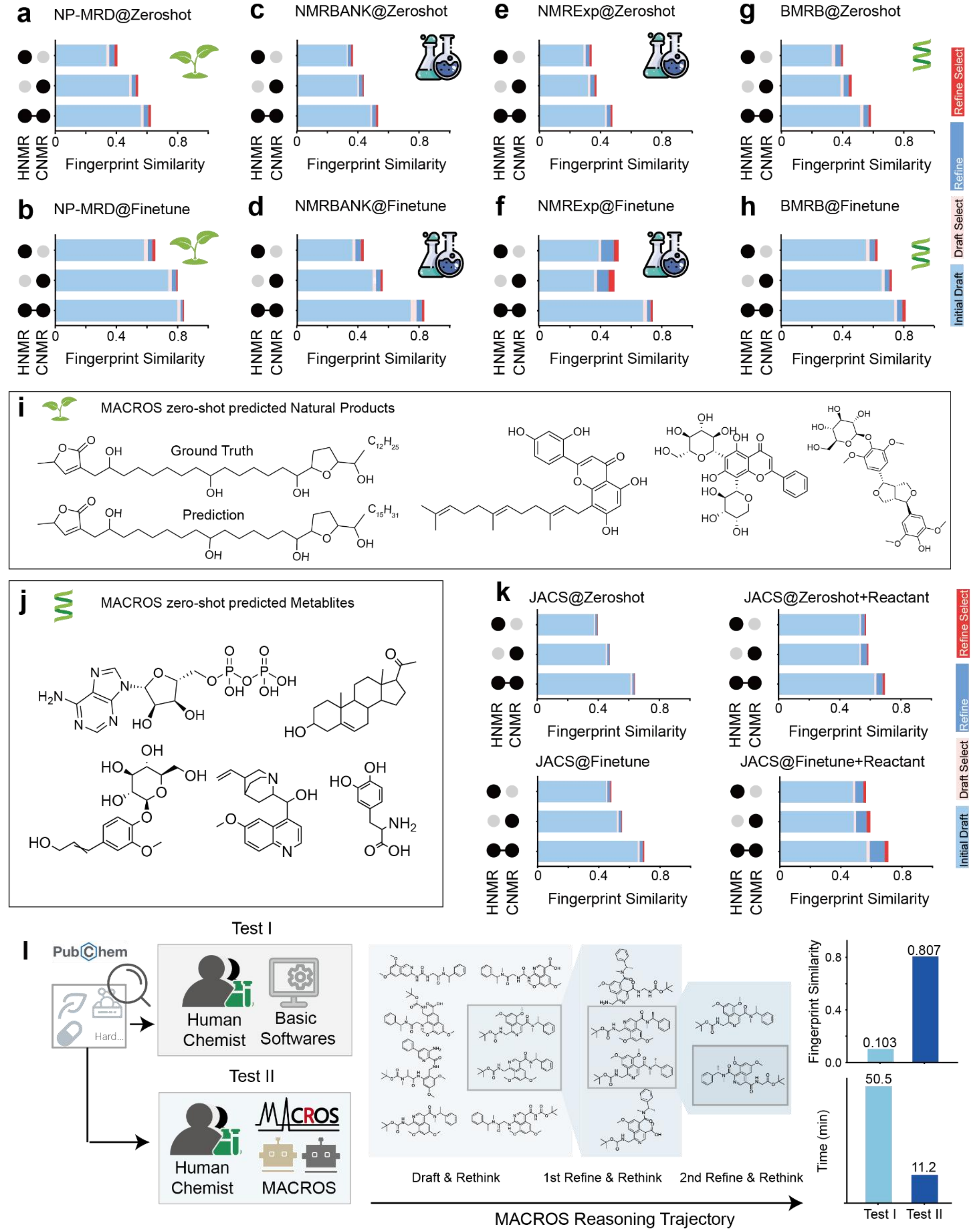

**Figure 4. Real-world validations and applications.**

**a-h**. Real-world zero-shot and fine-tuned performance of the molecular generation model evaluated on diverse NMR-based validation sources (a-b. natural products, c-f chemical reactions and g-h metabolites) using multiple spectral data combinations ($^1$H NMR only, $^{13}$C NMR only, and combined $^1$H/$^{13}$C NMR).

**i-j.** Representative examples of zero-shot parsing for natural products from (**i**) NP-MRD and (**j**) BMRB.

**k.** Zero-shot and fine-tuned annotation performance of chemical reactions.

**l.** Schematic diagram of reader studies: chemists vs. chemists + MACROS parsing distinct spectra collected from PubChem. The reasoning trajectory in MACROS significantly improves both parsing accuracy and efficiency.